\documentstyle[prb,aps,preprint]{revtex}

\def\5IAP{(5IAP)$_2$CuBr$_4 \bullet $2H$_2$O}
\def\...{$^{\ldots}$}
\def\o{$^{\circ}$}

\begin{document}
\draft
\preprint{}

\title{Magnetic properties of a new molecular-based spin-ladder system: 
\5IAP}
\author{C. P.  Landee$^1$, M. M. Turnbull$^2$, C. Galeriu$^1$, 
J. Giantsidis$^2$, and F. M. Woodward$^1$}
\address{$^1$Department of Physics, Clark University, Worcester, MA 01610 
\\ $^2$Carlson School of Chemistry and Biochemistry, Clark University, 
Worcester, MA 01610}
\date{\today}
\maketitle
\begin{abstract}
We have synthesized and characterized a new spin-1/2 Heisenberg 
antiferromagnetic ladder: bis 5-iodo-2-aminopyridinium tetrabromocuprate(II) 
dihydrate. X-ray diffraction studies show the structure of the compound to 
consist of well isolated stacked ladders and the interaction between the 
Cu$^{2+}$ atoms to be due to direct Br\...Br contacts. Magnetic 
susceptibility and magnetization studies show 
the compound to be in the strong-coupling limit, with
the interaction along the 
rungs ($J' \approx 13$\,K) much greater than the interaction along the 
rails ($J \approx 1$\,K). Magnetic critical fields are observed near 
8.3\,T and 10.4\,T, respectively, establishing the existence of the
energy gap. 
\end{abstract}


Quantum spins interacting antiferromagnetically on a lattice form one
of the simplest systems with which to explore cooperative effects in 
many body systems. The behavior can be particularly rich when the
parameters in the Hamiltonian can be tuned to a zero-temperature
quantum critical point \cite{sachdev00}. It is thought that 
high-temperature superconductivity exists due to its Hamiltonian's
proximity to such a point \cite{emery96,castellani96}. It has recently become
clear \cite{dagotto92,dagotto94,dagotto96} that a spin ladder forms
a remarkable system with which to explore quantum effects in
antiferromagnets. An even leg ladder has a cooperative singlet 
ground state, with a finite energy gap $\Delta$ in the spin
excitation spectrum. By application of a magnetic field, the size
of the gap can be reduced until it disappears at $H_{c1}$, at 
which point the system is at a quantum critical point. In
addition, holes injected into an even-leg ladder are predicted to pair
and possibly lead to superconductivity 
\cite{dagotto92,dagotto94,dagotto96}.

Few physical realizations of two-leg ladders are known, and only one
has been explored at the critical point. Among the most extensively
studied two-leg ladders are SrCu$_2$O$_3$ 
\cite{azuma94,johnston96,johnston00} and [(DT-TTF)$_2$][Au(mnt)$_2$] 
\cite{rovira97,arcon99}, but the strong exchange constants of these
compounds produce gaps too large to be overcome with available
magnetic fields. Only Cu$_2$(C$_5$H$_{12}$N$_2$)$_2$Cl$_4$ 
has exchange strengths small
enough to produce an experimentally accessible critical point,
$\mu_o H_{c1} = 7.5$\,T 
\cite{hayward96,chaboussant97b,weihong97,chaboussant97,hammar98,gu99}. 

We report here on the second two-leg spin ladder with an accessible
critical point at $\mu_o H_{c1} = 8.3$\,T. This new material, 
bis 5-iodo-2-aminopyridinium 
tetrabromocuprate(II) dihydrate (\5IAP), is the one of four new
spin ladders we have developed through the use of molecular-based
magnetism \cite{kahn93} and the first to be fully characterized by
high field measurements. The structure of the new compound will
readily permit dilution of the magnetic lattice, by which the
stability of the critical point can be explored further.

The structure of \5IAP consists of twin molecular chains 
cross-linked to form a ladder. The spin-ladders are well isolated 
from one another by the bulk of the organic molecule. As determined by 
X-ray diffraction, the crystallographic structure of \5IAP is triclinic, 
space group P\=1, $a = 6.7505$\,\AA, $b = 9.8433$\,\AA, 
$c = 16.7581$\,\AA, 
$\alpha = 78.512$\o, $\beta = 84.962$\o, $\gamma = 88.512$\o. 

The exchange interaction between the Cu$^{2+}$ ions takes place through 
Br\...Br direct contacts \cite{brown77,patyal90,zhou91}, as shown 
by dashed and dotted lines in 
Fig.\ \ref{ladder}. The interaction strength drops rapidly with the 
Br\...Br distance $r$, typically as $r^{-10}$, and depends also on the 
geometry of the superexchange pathway through such parameters as the mean 
trans angle $\theta$ of the distorted CuBr$_4$ tetrahedra, and the 
dihedral angle $\tau$ between the two Cu-Br\...Br planes 
\cite{straatman84,halvorson88}. 
The mean trans angle of the CuBr$_4$ tetrahedra in our system is 
$\theta = 132.7$\o. The exchange pathway along the rungs of the ladder 
is characterized by a distance between the Br atoms $r = 3.58$\,\AA, and 
by a dihedral angle $\tau = 180$\o. Along the rails the distance between 
the Br atoms is $r = 4.23$\,\AA, and the dihedral angle is $\tau = 
98.0$\o. The next shortest distance in the system between Br atoms is 
5.74\,\AA, along the diagonal of the ladder, too large to allow for an 
interaction. 

The model Hamiltonian for a spin ladder consisting of two coupled 
spin-1/2 Heisenberg chains is
\begin{eqnarray}
{\cal H} =&& J \sum_a \sum_i \bbox{S}_{i,a} \cdot \bbox{S}_{i+1,a} + 
J' \sum_i \bbox{S}_{i,1} \cdot \bbox{S}_{i,2} \nonumber \\ 
&&- g \mu_B H \sum_a \sum_i S_{i,a}^z,
\end{eqnarray}
where $\bbox{S}_{i,a}$ is the spin operator at site $i$ (the index $i$ 
runs along the chains) on the rail $a$ ($a$ = 1,2). $H$ is an external 
magnetic field in the $z$ direction. The exchange constants $J$ (along 
the rails) and $J'$ (along the rungs) are positive, corresponding to 
antiferromagnetic coupling. 

\begin{figure}
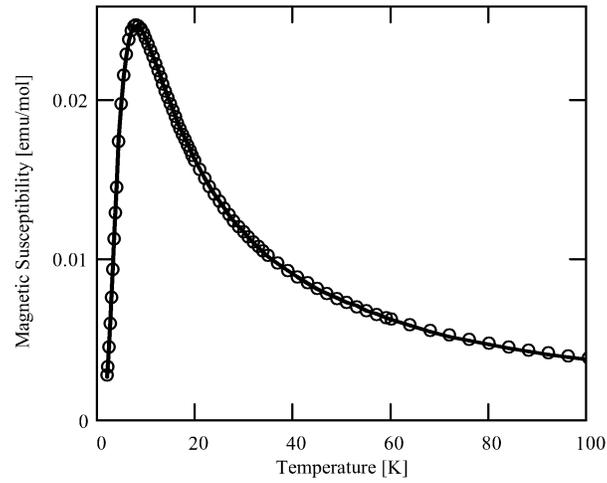

\caption{Temperature dependence of the powder dc magnetic 
susceptibility for \5IAP. The solid line is a fit (5) to the 
susceptibility of a spin-ladder plus a paramagnetic contribution.}
\label{chi}
\end{figure}

The powder dc magnetic susceptibility of \5IAP has been measured on a 
Quantum Design MPMS SQUID magnetometer, in a magnetic field of 1\,Tesla 
(T), from 2\,K up to 300\,K. The data have been corrected for the core 
diamagnetic (-379*10$^{-6}$\,emu/mol) and temperature independent 
paramagnetic (60*10$^{-6}$\,emu/mol) contributions. The magnetic 
susceptibility, plotted in Fig.\ \ref{chi}, shows a rounded peak 
around 8\,K, characteristic of low dimensional antiferromagnetic 
compounds, and a sharp drop at very low temperatures, characteristic 
to gapped systems. There is no sign of a structural phase transition 
at low temperatures. Fitting a Curie-Weiss law [$\chi = C/(T - \Theta)$] 
to high temperature data ($T \ge 20$\,K) gives 
$\Theta = - 5.3$\,K and $C = 0.416$\,emu\,K/mol. The 
sign of $\Theta$ confirms antiferromagnetic coupling between the spins. 
The Curie constant predicted from powder EPR measurements ($g = 2.10$) 
is 0.412\,emu\,K/mol, in good agreement.

Susceptibility data have been initially fitted using a high temperature 
expansion series for a spin ladder system \cite{weihong97}. Because it was 
determined in this way that the exchange interaction $J'$ along the 
rungs is 
much greater than the exchange interaction $J$ along the rails, we 
have been 
able to fit the data over the full temperature range by using an 
analytical formula resulted from third order perturbation theory 
\cite{johnston00}. 
Gu, Yu and Shen \cite{gu99} considered a dimer chain, in which the spin 
dimers are only weakly coupled, and calculated the perturbative 
corrections to the free energy due to the interactions of the dimers 
along the chain. From this expression for the free energy one can 
directly obtain the molar magnetic susceptibility in zero field (2). 
This expression reproduces with very high accuracy the QMC simulations 
of Johnston {\it et al} \cite{johnston00}, as long as the ratio $J/J'$ 
does not exceed 0.1.
\begin{eqnarray}
\chi_(T) =&& {4 C \over T} \Biggl[ {1 \over 3 + e^{2 \beta}} - 
{J \over J'} \Bigl[ { 2 \beta \over (3 + e^{\beta})^2} \Bigr] \nonumber 
\\ &&- \Bigl( {J \over J'} \Bigr)^2 \Bigl[ {3 \beta (e^{2 \beta} - 1) 
- \beta^2 (5 + e^{2 \beta}) \over 4 (3 + e^{\beta})^3} \Bigr] \nonumber 
\\ &&- \Bigl( {J \over J'} \Bigr)^3 \Bigl[ {3 \beta (e^{2 \beta} - 1) 
\over 8 (3 + e^{\beta})^3} \nonumber \\ &&- {9 \beta^2 e^{\beta} 
(1 + 3 e^{\beta}) - \beta^3 (7 e^{2 \beta} - 9 e^{\beta} -12) \over 12 
(3 + e^{\beta})^4} \Bigr] \Biggr],
\end{eqnarray}
where $C$ is the Curie constant for $Cu^{2+}$ ions
\begin{equation}
C = {N g^2 \mu_B^2 \over 4 k_B},
\end{equation}
and $\beta$ is the reduced temperature
\begin{equation}
\beta = {k_B T \over J'}.
\end{equation}

In addition to the expression (1) we have also allowed for a small 
paramagnetic impurity, to be determined from the fit: 
\begin{equation}
\chi_{total}(T) = (1 - x) \chi_(T) + x \chi_{imp},
\end{equation}
where $x$ is the concentration of the paramagnetic Cu$^{2+}$ ions. 
The exchange constants resulted from fitting our experimental data in 
this way are $J' = 12.95$\,K and $J = 0.75$\,K. The small paramagnetic 
component can be attributed to a contribution from 0.62\% of Cu$^{2+}$ 
ions due to lattice imperfections or impurity phases. The accuracy of 
the estimate of $J'$ and $J$ is affected by the strong correlation of 
the exchange parameters to the Curie constant \cite{johnston00}.

\begin{figure}
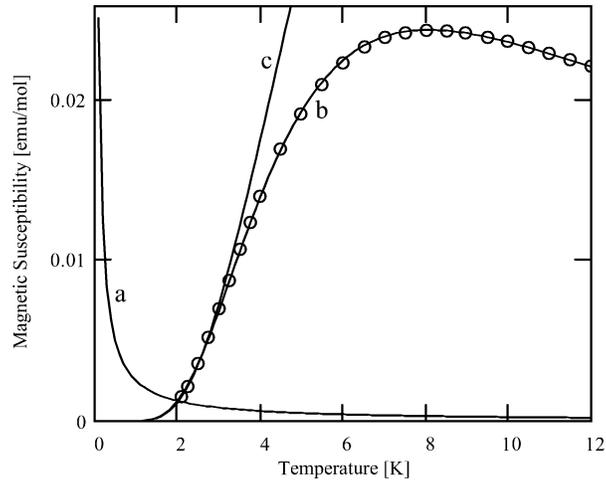

\caption{Low temperature behavior of the powder dc magnetic 
susceptibility for \5IAP. The circles are the experimental data 
from which we have subtracted the paramagnetic impurity contribution. 
This paramagnetic contribution is plotted as line (a). Line (b) is 
the perturbative expression (2), and line (c) is the low temperature 
expression (7).}
\label{gap}
\end{figure}

The intrinsic part of the susceptibility $\chi$ of our dimer-like ladder 
system is estimated by subtracting the paramagnetic component, and is 
plotted in Fig.\ \ref{gap} with open circles. The abrupt decrease of 
$\chi$ toward zero clearly suggests the presence of an energy gap in the 
spin excitation spectrum. This gap, for $J \ll J'$, can be directly 
evaluated from a strong-coupling series expansion \cite{cabra97}
\begin{equation}
\Delta = J' - J + {J^2 \over 2 J'} + {J^3 \over 4 J'^2} - 
{J^4 \over 8 J'^3} + {\cal O}(J^5);
\end{equation}
for our system $\Delta$ = 12.23 K.

According to a theoretical study of the two-leg Heisenberg ladder 
system \cite{troyer94}, if the continuum of the first excited states 
has a parabolic dispersion, and if magnon interactions are ignored, at 
low enough temperatures $T \ll \Delta$ the magnetic susceptibility as a 
function of temperature should be
\begin{equation}
\chi(T) = {\alpha \over \sqrt{T}} exp(-{\Delta \over T}),
\end{equation}
where $\alpha$ is a constant corresponding to the dispersion of the 
excitation energy. As seen in Fig.\ \ref{gap} the solid line (c) 
calculated using $\Delta = 12.23$\,K and $\alpha = 0.746$\,K$^{1/2}$ 
emu/mol reproduces the experimental data rather well, and is 
indistinguishable from the curve given by (2) for temperatures up to 
3.75\,K, corresponding to $T/\Delta \le 0.3$. Equation (7) can fit the 
susceptibility data for even higher temperatures, but the energy gap 
is not accurately obtained in this way \cite{azuma94}. 

\begin{figure}
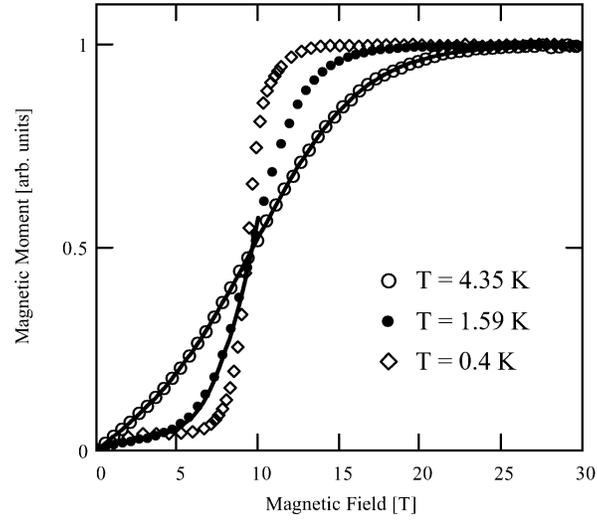

\caption{High field magnetization curve for different temperatures for 
\5IAP. $T = 0.4$\,K for the diamonds, $T = 1.59$\,K for the filled 
circles, and $T = 4.35$\,K for 
the open 
circles. Only representative points of the full data sets have been 
plotted. The black lines are theoretical predictions for 
$J' = 12.99$\,K, $J = 1.15$\,K, and $x = 3.2$\%.}
\label{mag}
\end{figure}

Due to the weak exchange constants involved, we have been able to fully 
saturate the magnetization of our sample in high magnetic fields. These 
measurements have been performed at the 
two facilities of the 
National High Magnetic Fields Laboratory.
A sample of \5IAP 
was studied in Tallahassee in dc-fields up to 30\,T,
at temperatures of 4.35\,K and 1.59\,K, and in Los Alamos
in pulsed fields up to 50\,T, cooled down to 0.4\,K.
The data are shown
in Fig. \ \ref{mag}. 
At low fields and at the lowest temperatures, the 
spin-ladder system is in the non-magnetic singlet ground state,
and the small moment observed is due only to the
paramagnetic impurity phase. 
At the lower critical field $H_{c1}$ 
the gap 
is closed, and the magnetization increases almost linearly with 
field up to  the second critical field $H_{c2}$, where 
the system is saturated. 
At the higher temperature $T = 4.35$\,K the critical 
fields are less evident, due to thermal population of the excited 
states.

An expression for the magnetization of a ladder as a function of 
temperature and magnetic field has been obtained from the free energy 
\cite{gu99} previously mentioned. The perturbation series for the 
magnetization involves powers of $J/J'$ and $J/k_B T$. For good 
convergence of the series, we should have $J << k_B T$. The magnetic 
field enters the magnetization expression as the exponent 
$n g \mu_B B/k_B T$, where $n$ is an integer, increasing with 
the order of perturbation. Consequently, the calculated expression 
of the magnetization departs from the experimental data for low 
temperatures ($T = 1.59$\,K) and high magnetic fields ($B > 10$\,T), due 
to poor convergence of the series. For the data set at the higher 
temperature $T = 4.35$\,K, the calculated and the experimental 
magnetization agree well for the full magnetic field range.

We need only three parameters to reproduce the reduced 
magnetization $M/M_{sat}$. These parameters are $J$, $J'$, and the 
percentage of paramagnetic spins. Because a 1/2 spin ladder has 
a gap in the energy spectrum, at very low temperatures and magnetic 
fields the system will be in the singlet ground state with no 
magnetic moment. The only contribution to the magnetization will 
come from the paramagnetic impurities. This contribution is clearly 
identified for the $T = 0.4$\,K and $T = 1.59$\,K data at low fields. 
The two exchange 
constants and the impurity percentage have been extracted from a fit of 
the $T = 4.35$\,K data to the calculated magnetization, 
and the values obtained are $J' = 12.99$\,K and $J = 1.15$\,K, 
corresponding to a gap of $\Delta = 11.90$\,K. The concentration of 
the paramagnetic spins extracted from the magnetization fit is 
3.2\% of the Cu$^{2+}$ ions. The higher value is justified by the 
fact that we have used a different sample than in the susceptibility 
measurements. The purer batch has been obtained by a slower growth, 
minimizing the concentration of impurities.

The magnetization fit, unlike the susceptibility fit, is not affected 
by uncertainties in the nature of the impurities. This is most clearly 
seen from the analysis of the two critical fields $H_{c1}$ and $H_{c2}$ 
\cite{chaboussant97b,chaboussant98}. At the lowest temperature (0.4\,K) 
the magnetization of the spin ladder is close to zero up to the lower 
critical field $H_{c1}$ where the energy gap is closed by the magnetic 
field. If one assumes that the first excited states are triplets, then 
$\mu_o H_{c1} = \Delta/(g \mu_B) \approx (J' - J)/(g \mu_B) = 8.39$\,T. 
The 
ground state is fully aligned above the higher critical field 
$\mu_o H_{c2} = 
(J' + 2 J)/(g \mu_B) = 10.83$\,T, where the magnetization reaches its 
saturation value. The critical fields determined from the $T = 0.4$\,K 
data are $\mu_o H_{c1} \approx 8.3$\,T and 
$\mu_o H_{c2} \approx 10.4$\,T. The 
magnetization increases almost linearly between $H_{c1}$ and $H_{c2}$. 
The exchange constants are unambiguously linked to the position and slope 
of this domain in which the magnetization increases almost linearly.

In summary, \5IAP has been shown to behave as a $S = 1/2$ 
antiferromagnetic ladder with a dominant rung interaction $J' = 
13.0(1)$\,K and a rail interaction $J = 1.0(2)$\,K. The ratio $J/J' = 
(7.7 \pm 1.5)$\% is about 40\% of the value found for 
Cu$_2$(C$_5$H$_{12}$N$_2$)$_2$Cl$_4$ \cite{chaboussant97b}. The two 
critical fields near 8.3\,T and 10.4\,T are within the range of common 
superconducting magnets, and hence it will be possible to examine the 
behavior 
of this quantum system when the gap has been closed ($H > H_{c1}$), as 
has been recently done for Cu$_2$(C$_5$H$_{12}$N$_2$)$_2$Cl$_4$ 
\cite{chaboussant98,chaboussant00,calemczuk99}.

The high magnetic field measurements 
were performed at the 
National High Magnetic Fields Laboratory in
Tallahassee and Los Alamos. 
The NHMFL is supported by the NSF cooperative agreement No. 
DMR-9527035 and the State of Florida. We acknowledge the technical
assistance of Donavan Hall and Neil Harrison. The Clark University
research group received support from the NSF under grant No.
DMR-9803813.

\begin{figure}
\caption{Crystal packing of the Cu$^{2+}$ metal ions, along with the 
path of exchange interaction. The Cu$^{2+}$ ions are the small circles, 
the Br$^-$ ions are the larger circles. 
The Br-Br contacts are 3.58\,\AA 
(rung) and 4.23\,\AA (rail) with the rails aligned vertically 
along the page. The organic molecules are not shown.}
\label{ladder}
\end{figure}

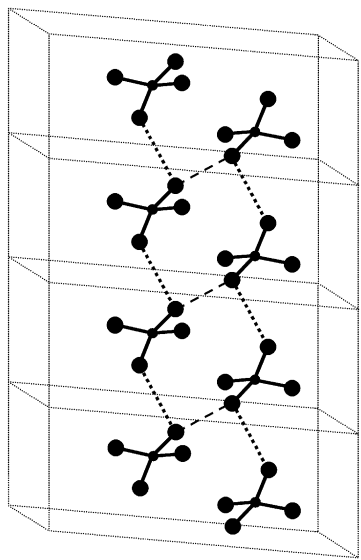
\end{document}